\documentclass[showpacs,amsmath,amssymb,amsthm,amsfonts,11pt]{revtex4}

\usepackage{natbib}
\usepackage{graphicx}
\DeclareGraphicsExtensions{.jpg,.jpeg,.pdf,.png,.mps,.eps,.ps,.gif}
\usepackage{color}
\newcommand{\bminil}[1]{\begin{minipage}[l]{#1 \textwidth}}
\newcommand{\bminir}[1]{\begin{minipage}[r]{#1 \textwidth}}
\newcommand{\bminic}[1]{\begin{minipage}[c]{#1 \textwidth}}
\newcommand{\emini}{\end{minipage}}
\newcommand{\EQL}{\begin{equation}\label}
\newcommand{\EQ}{\begin{equation}}
\newcommand{\EN}{\end{equation}}
\newcommand{\BFG}{\begin{figure}}
\newcommand{\EFG}{\end{figure}}
\newcommand{\ITM}{\begin{itemize}}
\newcommand{\ITN}{\end{itemize}}
\newcommand{\p}{\partial}

\newcommand{\half}{\mbox{$\frac{1}{2}$}}

\newcommand{\ddt}{\frac{d}{dt}}

\newcommand{\dddt}{{\displaystyle\frac{d}{d t}}}

\newcommand{\bS}{\mbox{\boldmath S}} 
\newcommand{\bu}{\mbox{\boldmath u}} 
\newcommand{\authone}[2]{#2 #1.}
\newcommand{\authtwo}[4]{#2 #1, #4 #3.}
\newcommand{\auththr}[6]{#2 #1, #4 #3, #6 #5.}

\newcommand{\yjour}[8]{ #6~{\em #2}~#1;#3:#4#5.}
\newcommand{\yproc}[7]{ #7~In:~#2, editors. {\em #3}~#4 #1 #5#6.}

\begin{document}
\title{Bounds on a singular attractor in Euler using vorticity moments}
\author{Robert M. Kerr}
\affiliation{
Department of Mathematics, 
University of Warwick, Coventry CV4 7AL, United Kingdom}

\begin{abstract} 
A new rescaling of the vorticity moments and their growth terms is used 
to characterise the evolution of anti-parallel vortices governed by the 3D Euler 
equations. To suppress unphysical instabilities, the initial 
condition uses a balanced profile for the initial magnitude 
of vorticity along with a new algorithm for the initial vorticity direction. 
The new analysis uses a new adaptation to the Euler equations of 
a rescaling of the vorticity moments developed for Navier-Stokes analysis. 
All rescaled moments grow in time, with the lower-order moments bounding 
the higher-order moments from above, consistent with new results 
from several Navier-Stokes calculations. 
Furthermore, if, as an inviscid flow evolves, this ordering is assumed to hold, 
then a singular upper bound on the growth of these moments can be used to 
provide a prediction of power law growth to compare against. There is a significant 
period where the growth of the highest moments 
converges to these singular bounds, demonstrating a tie between the strongest 
nonlinear growth and how the rescaled vorticity moments are ordered. 
The logarithmic growth of all the moments are calculated directly and 
the estimated singular times for the different $D_m$ converge to a 
common value for the simulation in the best domain.


\end{abstract}
\maketitle

\email{Robert.Kerr@warwick.ac.uk}

\begin{center} To appear in the {\it Procedia IUTAM} volume of papers for
{Topological Fluid Dynamics II} under \\
The growth of vorticity moments in the Euler equations \end{center}

\section{Background}

Two unresolved issues that have limited the application of numerics to 
the vortex dynamics and regularity questions of the three-dimensional Euler equations 
have been the inadequate analysis tools and the difficulties in specifying 
reproducible initial conditions. 
The existing analysis tools are unable to simultaneously cover the 
necessary range of scales in both space and time, while existing methods for mapping 
vortex tubes onto Eulerian 
meshes tend to generate ghost images unless {\it ad hoc} massaging is applied. 
This has led to weak and conflicting conclusions that depend 
upon the numerical method used and the choice of analysis that 
is applied to the results.
 
To address these problems, this paper introduces an improved initialization 
for curved vortex tubes following an arbitrary trajectory and new analysis 
that is 
based upon higher-order vorticity moments, 
and then applies these to simulations of interacting anti-parallel vortices. 
The new initialisation suppresses core instabilities, which eliminates 
the ghost vortices found in earlier work \cite{HouLi06} and discussed by \cite{BustaKerr08}. 
Furthermore, the new trajectory algorithm allows the evolution of vortices with the same local perturbation, 
but different lengths, to be compared.  One example calculation is given in Fig. 1. 
The new analysis allows one to compare all orders of the vorticity moments and 
generates new bounds against which to compare the results, 
which leads to more robust conclusions.

The new analysis tool is an adaptation to the inviscid Euler equations of a new 
rescaling of the vorticity moments for the viscous Navier-Stokes equations. 
The rescaling uses a new frequency $\varpi_0$, plus scaling powers $\alpha_m$,
to convert the standard $\Omega_m$, or $L^{2m}$, vorticity moments, 
into the following $D_m$ moments \cite{Gibbon10,Gibbon12}: 
\EQL{eq:Dm} D_m=(\varpi_0^{-1}\Omega_m)^{\alpha_m}\quad{\rm where}\quad
\Omega_m=\left(L^{-3}\int_{\cal V}|\omega|^{2m}dV\right)^{1/2m}
,~~ \varpi_0=\varpi_\nu=\nu/L^2~~{\rm and}~~\alpha_m=2m/(4m-3)\,.\quad \EN
For the Navier-Stokes equations, $\varpi_0$ is based upon the 
viscosity $\nu$ and the characteristic large length scale $L$ of the turbulence and 
the $\alpha_m$ are designed such that neighbouring 
$D_m(t)$ and $D_{m+1}(t)$ terms can be compared directly using 
Navier-Stokes vorticity moment inequalities. 
This is adapted to the inviscid case below.
The new rescaling makes comparisons between all
the moments of the vorticity possible, both analytically and numerically.

Historically, only the two limiting $D_m$ have been used for addressing 
regularity questions. These are the global mean square vorticity or enstrophy, 
rescaled here into $D_1$, and the point-wise maximum of vorticity 
$\|\omega\|_\infty$, rescaled into $D_\infty$. This is in part because analysis 
of the inequalities relating the intermediate moments had never been done. 
The known importance of $D_1$ is for addressing Navier-Stokes regularity 
(see references in \cite{Doering09}), 
while for the Euler equations, possible singularities are controlled by the 
time integral of $\|\omega\|_\infty$. That is, for the Euler equations, if
\EQL{eq:BKM} \int_0^t \|\omega\|_\infty d\tau < \infty
\quad\mbox{for all time}~t>0\,,\EN
then the Euler equations are regular \cite{BKM84}.
\index[authors]{Beale, J.T.}

The importance of the $D_m$ between these limits is that by taking
their ratios, new criteria for the regularity of the Navier-Stokes equations 
can be found \cite{Gibbon12b}. To demonstrate the usefulness of the Navier-Stokes 
$D_m(t)$, Fig. 2(left) shows their evolution using data from 
a viscous, anti-parallel reconnection calculation using the initial 
condition described below. A full discussion of the trends, with lower order 
bounding higher order and convergence as $m$ increases, is being prepared for
publication.

To adapt this rescaling to vorticity moments of the inviscid Euler equations 
a non-viscous replacement for the scaling frequency $\varpi_0$ in \eqref{eq:Dm} 
is needed.  The inviscid modification chosen here defines $\varpi_0$ using 
the circulation of the vortices $\Gamma$ instead of the viscosity $\nu$. 
That is: $\varpi_0=\varpi_\Gamma=\Gamma/L^2$.  

For $m<\infty$, a computational advantage of using these inviscid $D_m$ 
in numerical analysis of the Euler equations is that they and their 
time derivatives $dD_m/dt$ 
can be determined at run-time and then compared as functions of time 
to integrals suggested by mathematical analysis. Furthermore, from 
the inverses of the logarithmic time derivatives
$(d\log(D_m)/dt)^{-1} = D_m/(dD_m/dt)$, one can estimate the type of 
power-law singular growth using simple time differences \cite{BustaKerr08} 
or, if it is assumed that the $D_m^{-2}\rightarrow a(T_m-t)$, 
running estimates of the $T_m(t)$ can be made without using time 
differences. These running estimates will be used in the final test for 
singular growth using data from the best of the new anti-parallel Euler calculations.

When these new Euler simulations were begun, the modest goal was to explain the 
type and strength of the convergence of the $D_m$ moments in an early period of 
the Navier-Stokes calculation. The desired comparison period would be up to the 
beginning of the first vortex reconnection event at $t=16$, 
shown in Fig. \ref{fig:T0-16}(right). Before $t=16$, the viscous effects in the 
Navier-Stokes calculation should be negligible and the nonlinear terms, 
shared with the Euler equations, should dominate. 
The two frames in Fig. \ref{fig:DmNSEuler} are used to compare these 
$D_m$ trends for the two Navier-Stokes and Euler calculations being 
highlighted here.

The observed ordering of the $D_m(t)$ for the Navier-Stokes simulations in 
Fig. \ref{fig:DmNSEuler}(left), 
plus the period of extended singular growth of the $D_m^{-2}$  
Euler moments in Fig. \ref{fig:DmNSEuler}(right) 
then led to Fig. \ref{fig:Dm-2scaling}. This figure addresses the question 
of whether the Euler equations have a singularity \index[subject]{singularity}
using a new set of numerically determined time integrals and analysis of 
logarithmic time derivatives found at run-time. 
The new time integrals come from mathematical analysis of the Euler equations 
that assumes {\it a priori} 
that the order of the $D_m$ seen numerically will hold for all time, 
as indicated by Fig. \ref{fig:DmNSEuler}(right).

All the calculations are, fundamentally, in periodic computational domains, 
with symmetries used to decrease the data and time needed to do the 
calculations. Several filtered/dealiased pseudospectral methods have been 
tested and described previously \cite{BustaKerr08}. 
The method chosen for the calculations here is a 
combination of the 2/3rds dealiasing rule plus a 36th order filtering 
method that was first introduced without dealiasing \citep{HouLi06}. 
\index[authors]{Kerr, R.M.}
\index[authors]{Bustamante, M.D.}
\index[authors]{Hou, T.Y.}

The principle axes are: 
$x$ is the direction of propagation of the vortex pair, $y$ is in the 
primary direction of the vortices, and $z$ is the direction between the 
vortices. The computed domain size is $L_x\times L_y\times L_z$, while 
the fully-periodic domain would be in $L_x\times 2L_y\times 2L_z$.
Domain sizes and meshes are given in Table 1. 
Referring to the initial condition in Fig. \ref{fig:T0-16}, 
the $y=0$ symmetry plane with the maximum perturbation will be called the 
{\it perturbation plane} and the $z=0$ symmetry plane between the vortices will 
be called the {\it dividing plane}.

The paper is organised as follows. First, the new initial condition 
is described briefly. Next, the re-scaling of the vorticity moments for 
the Navier-Stokes and Euler equations is discussed further and applied to 
the new calculations, from which a new ordering for the $D_m(t)$ moments 
is found. This ordering has been found for all times for both the viscous and 
inviscid cases. After the new ordering is established, 
new upper bounds on the growth of the moments in the Euler equations 
are found and applied to the inviscid Euler solutions. 
Next, the logarithmic time derivatives of the $D^2_m$ from the Euler 
calculation are used to give running estimates of the singular times, labeled 
$T_m(t)$. It is found that 
for $m>1$, these estimated times converge. That is, all the 
$T_m(t)\rightarrow T_c$. Finally, there is some discussion of additional 
diagnostics that are now being collected, such as the curvature of the vortex lines, 
that will be needed if we are going to understand why
the Euler equations can obey singular scaling laws for extended periods.

\section{Initial condition}

At meeting on the Euler equations in 2007 in Aussois, France, one topic was
results from direct numerical simulations that addressed the question of
regularity of the Euler equations.
The conclusions of the two anti-parallel calculations \cite{BustaKerr08,Hou08} were 
different, even though both were nominally using initial conditions similar 
to \cite{Kerr93}. 
\index[authors]{Kerr, R.M.}
\index[authors]{Bustamante, M.D.}
\index[authors]{Hou, T.Y.}
Clearly, the prescription in \cite{Kerr93} was flawed.
These flaws have now been identified and
will be described in detail in another paper.  The three primary elements
of the new initial condition are these:
\ITM
\item A new profile of the vorticity distribution in the core that is
based upon the Rosenhead regularisation of a 2D point vortex and
is similar to the two-dimensional density
profiles used for quantum Gross-Pitaevskii calculations \cite{Kerr11}.
\item A new direction algorithm that, for a given $(x_i,y_j,z_k)$ on the 
three-dimensional grid, begins by finding the nearest position $(x_s,y_s,z_s)$ 
on the given analytic trajectory.  The distance used in the profile function for
finding $|\omega|(x_i,y_j,z_k)$ is $r=|(x_i,y_j,z_k)-(x_s,y_s,z_s)|$ and
the direction of the vorticity at the points $(x_i,y_j,z_k)$ is given by
the tangent of the chosen trajectory at $(x_s,y_s,z_s)$.
\item Making the vortices very, very long to minimise boundary effects.
\ITN 

The resulting profile has been used for anti-parallel vertical vortices in 
a stratified fluid, the anti-parallel unstratified Navier-Stokes calculations mentioned 
here, and now anti-parallel Euler vortices. In each case, unphysical initial 
instabilities due to small-scale inbalances have been suppressed, 
a cleaner and stronger 
larger-scale instability has been identified, and, where appropriate, a 
transition to sustained turbulence forms from the vortex interactions where none 
had been seen in earlier work. Unlike in earlier work \cite{BustaKerr08,Kerr93}, 
no extra massaging or squeezing of the initial condition is 
needed to ensure that there is only one sign of the vorticity in the calculated 
$y=0$ perturbation plane.

The computational procedure is as follows: First, the vortex is initialised 
on a modest mesh, which is then put onto a much larger computational mesh 
by adding zeros at 
the higher wavenumbers. The calculation then proceeds on this mesh until, 
by comparing results on different meshes, the collapse has progressed to the 
point where the calculation would soon be underresolved. Then the 
calculation is remeshed onto a finer mesh. Two remeshings are typically 
needed to reach the final mesh at the final times. The calculations 
used for the current study are given in the Table 1.

\begin{figure} 
\begin{center}
\includegraphics[width= 8 cm]{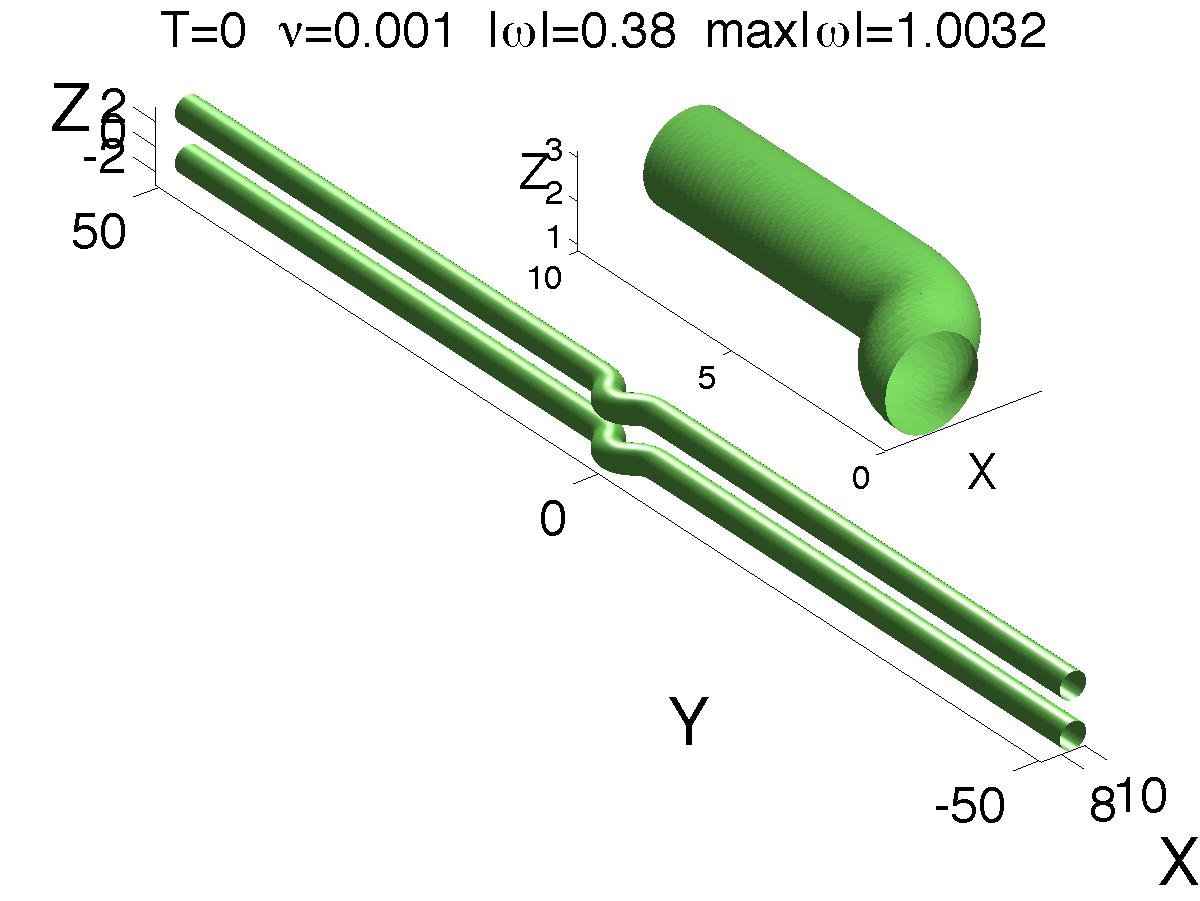}\hspace*{5mm}
\includegraphics[width= 8 cm]{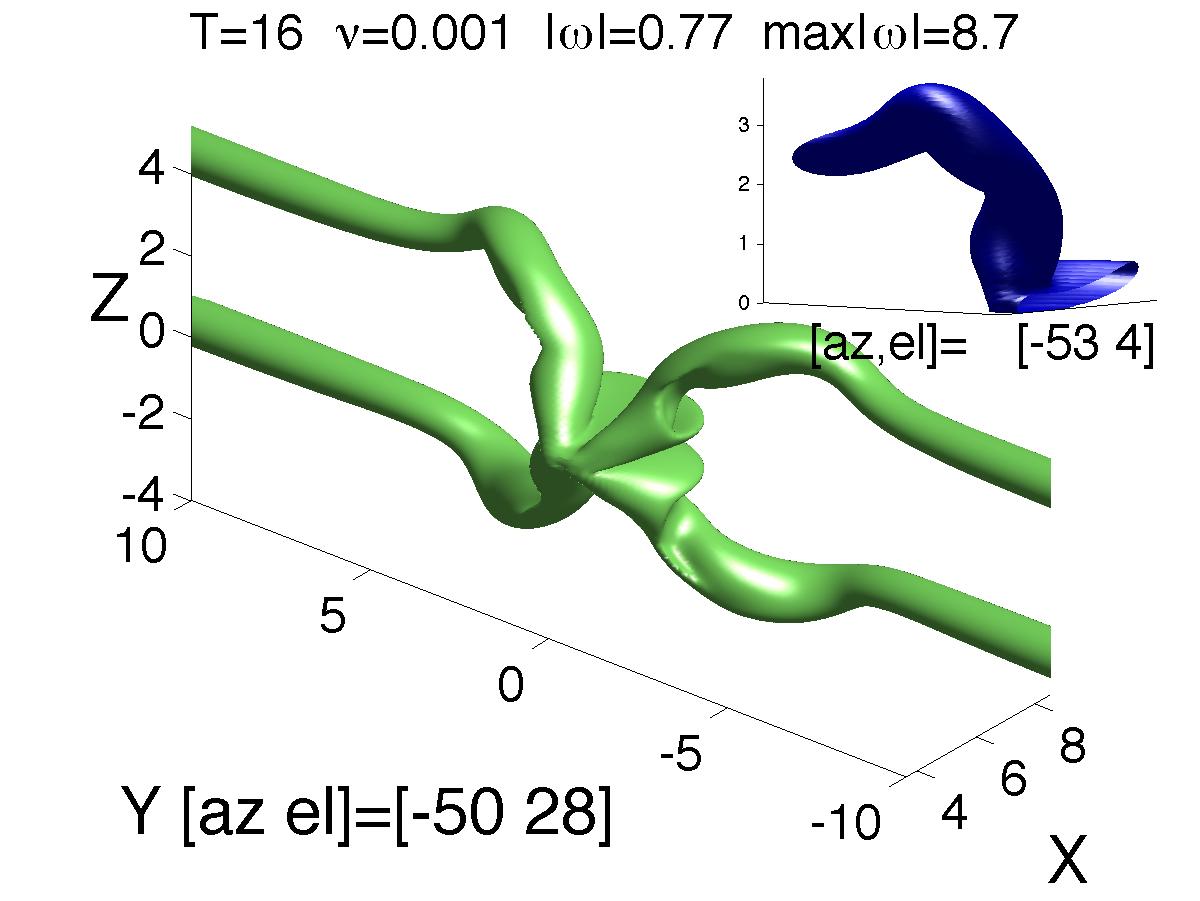}
\vskip6pt
\caption{\label{fig:T0-16} {\bf Left:} Very long, anti-parallel initial
condition at $t=0$. {\bf Right:} $t=16$ (Navier-Stokes).
Insets show $z>0$ for $0\leq y\leq10$.}
\end{center}
\end{figure}

The initial and evolved Navier-Stokes vorticity isosurfaces in 
Fig.  \ref{fig:T0-16} apply to both the Navier-Stokes and Euler calculations 
because viscous dissipation for the Navier-Stokes case at $t=16$ has been minimal.
The insets show the upper/left quarter domain near the $y=0$ perturbation plane,
with the $t=0$ inset showing that the initial vortex tube has a circular cross-section
of  constant width along its entire length.

These figures can be compared with similar stages in the evolution of
anti-parallel quantum vortices in \cite{Kerr11} and to the cover illustration
in \cite{Kerr96}, which shows how the vortex lines twist as they extend from
the $y=0$ perturbation plane.  The vortices do more than twist.  They
actually bend back upon themselves until, near $y=\pm 7$, they are
closer than the unperturbed original vortices for $y>\pm 8$.  The possible
significance of this bend and its curvatue will be discussed in the summary.

\section{Navier-Stokes intermittency and the rescaling vorticity moments}
\index[subject]{vorticity moments}

A neglected topic in studying the Navier-Stokes equations
is temporal intermittency, periods of intense activity, interspersed by 
relatively quiescent periods. One approach to characterising this type
of intermittency is through higher-order strain
($[\bS]=S_{ij}=0.5(\p u_i/\p x_j+\p u_j/\p x_i)$) and vorticity
($\omega=\nabla\times\bu$) moments, plus experimentally measurable single-point 
derivatives \cite{SreeniAntoniaAR97}.  Numerically, convergent statistics for
$\overline{\bS^{2m}}$ and $\overline{\omega^{2m}}$ with $m$=2 and 3 
were obtained as early as 1985 \cite{Kerr85}.

However, having only orders $m$=2 and 3 is insufficient for making theoretical
comparisons  and it has been impossible to get convergent statistics for 
the next higher-order moments for even the largest forced simulations
\cite{IshiAR09}.  The problem is two-fold.  First, the difference between the
higher-moments moments in the quiet periods and the intense periods can
be huge, and second, these occur on the time-scale of the large-scale forcing
for simulations that, due to their size, can only be run for a few
of these characteristic timescales. 

Recently, Yeung, Donzis \& Sreenivasan \cite{YeungDonzisSreeni12} 
\index[authors]{Yeung, P.K.}
have found that convergent statistics for their forced simulations
can be obtained by taking ratios of the higher-order moments. 
While simultaneously, new mathematics
has concluded that these ratios, rescaled in a manner
consistent with inequalities for the time derivatives of the higher-order  
vorticity moments $D_m$ \cite{Gibbon10}, can give new insight into the
Navier-Stokes singularity question \cite{Gibbon12},
\index[authors]{Gibbon, J.D.}
as summarized in \cite{Kerr12}.

One analytic approach to answering whether the Navier-Stokes equations 
are regular or not starts by assuming that there are quiescent 
and intensely intermittent periods, called 
“good” and “bad”, or possibly “neutral” \cite{Gibbon10}. 
Because the $D_m(t)$ moments can be compared directly 
using vorticity moment inequalities, the new mathematics \cite{Gibbon12} 
is able to derive new bounds on the periods of the maximum growth of
the $D_m$ that can be compared to numerical results.

\begin{figure}
\begin{center}
\includegraphics[width=7 cm]{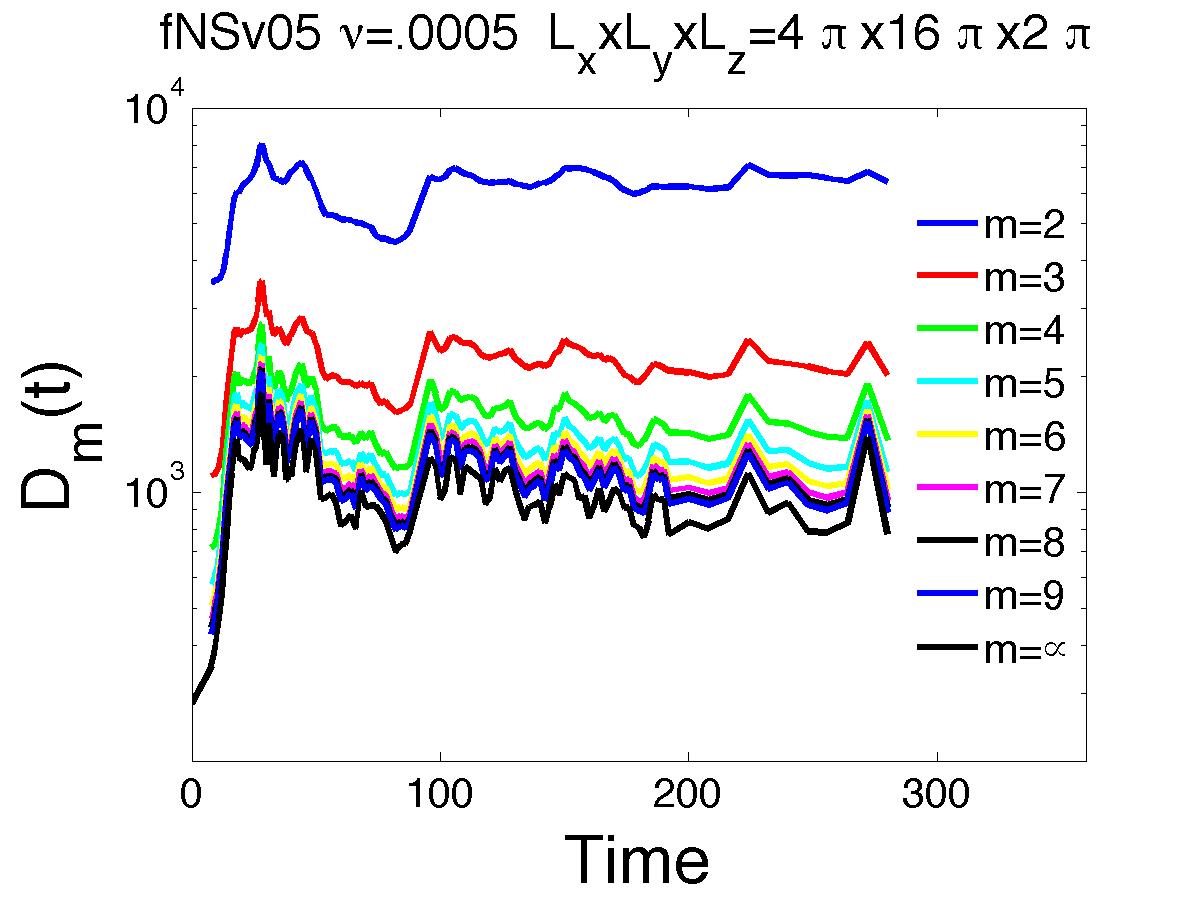}
\hspace*{5mm}
\includegraphics[width= 7 cm]{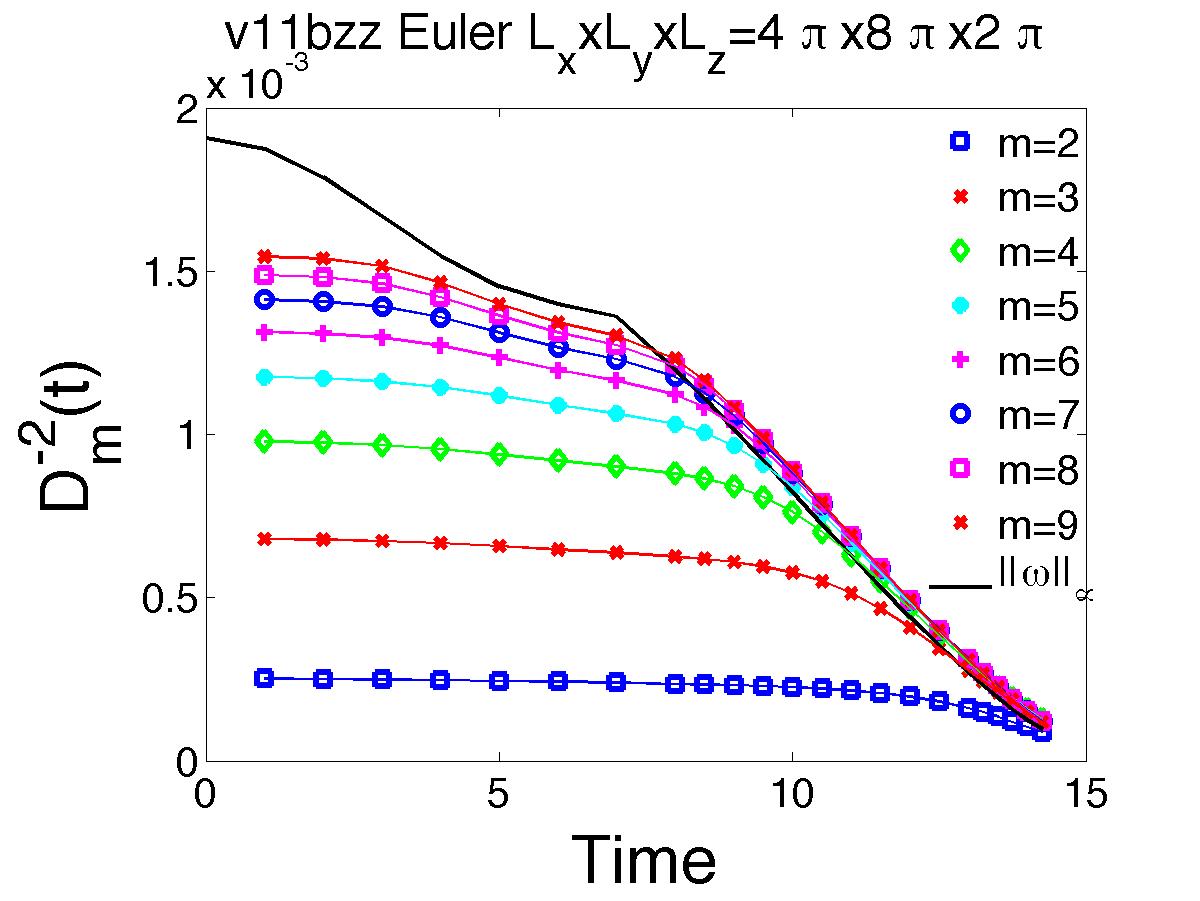}
\caption{\label{fig:DmNSEuler}
\footnotesize{{\bf Left:} $Re=\Gamma/\nu=4000$: $D_m$ \eqref{eq:Dm}
from an anti-parallel calculation. $D_m$ are ordered with lower-order
bounding higher-order for all times. Especially note the the periods of
steepest growth, $t<16$ and $t\approx90$ when the nonlinear terms dominate.
{\bf Right:} The inverses: $D_m^{-2}(t)$,
from the Euler calculation with a similar initial condition for the first
period of sharp growth ($t\leq 15$). The
hierarchy of $D_m^{-2}(t)$ includes $\varpi/\|\omega\|_\infty$. The $D_m^{-2}(t)$ 
from the Navier-Stokes calculation are similar, but with
a greater deviation from a linear form as time increases.
}}
\end{center}
\end{figure}

Fig. \ref{fig:DmNSEuler}(left) shows how the $D_m$ are ordered for the new
reconnection calculation, with the lower order
$D_m$ bounding the higher order $D_m$ for all times: $D_{m+1}(t)<D_m(t)$. This means
that the different $D_m$ never cross one-another and the definitions of
bad and good periods are the same for all of the $D_m$.
This ordering of lower-order above higher-order $D_m$ was unexpected because
it is opposite of how the original $\Omega_m$, without rescaling,
are required to be ordered using H\"older inequalities and is opposite
to what would easily ensure regularity of the Navier-Stokes equations
using the new bounds of \cite{Gibbon12}.
\index[authors]{Gibbon, J.D.}
This ordering has now been identified in every Navier-Stokes
simulation it has been tested against.  A joint paper is in preparation
and mentioned in \cite{Gibbon12b}.

What governs the dynamics during these intense/bad and quiet/good periods?
Fig. \ref{fig:T0-16}(right) shows the structures at the end of the most
extreme growth of the $D_m(t)$, up to $t=16$,
when the first Navier-Stokes vortex reconnection is forming. In a new
Navier-Stokes reconnection paper it will be shown that all of the
subsequent periods of intense growth of the higher-order $D_m$
can be tied to how vortices are attracted and stretched
just before reconnection events.  

\section{Rescaled vorticity moments and the Euler equations}

The strongest growth of the Navier-Stokes $D_m$ in Fig. \ref{fig:DmNSEuler}(left) 
is before the first reconnection at $t\approx16$, when viscous effects are negligible 
and the nonlinear Euler dynamics are strongest. During this period, the growth of the 
normalised enstrophy production, a skewness factor, is is up to three times 
the values typically determined in large Reynolds number experiments and simulations.
In order to understand the origins of this period of growth, a new series of 
simulations of the Euler equations were begun that 
cover part of this period ($t \leq14.25$). 
The particular calculation shown is just one of a series of new 
anti-parallel inviscid Euler calculations outlined in Table 1, all using the new 
profile and trajectory algorithms. The objective was to find, and confirm, whether 
a domain could be identified where the boundaries were not suppressing any growth 
of the maximum of vorticity $\|\omega\|_\infty$. And then, determine if this
calculation indicates singular behaviour, or not.

It was found that changing the length of the 
domain in the $y$-direction had the greatest effect upon the growth
of $\|\omega\|_\infty$, with its growth being suppressed until 
$L_y=8\pi$ was reached, the case labeled v11bzz. The v11g case with
$L_y=16\pi$ gave identical results to the v11bzz case.

The choice of $\varpi_0=\varpi_\Gamma$ for the Euler $D_m$ analysis
was in part inspired by how the growth of enstrophy in an 
earlier Euler calulation \cite{KerrBustamante11} could be explained 
empirically by replacing viscosity $\nu$ with the circulation $\Gamma$ in
the well-known inequality for the upper bound on enstrophy growth in
Navier-Stokes:: $(d/dt) \Omega_1^2\leq C_1 (\Omega_1^2/(\nu/L^2))^3$. 
This empirical guess can now be replaced by robust Euler bounds that use
the $D_m$ with $\varpi_0=\varpi_\Gamma$.

Following the proof of Proposition 1 \cite{Gibbon12b}, one starts with:
\EQL{eq:dtOm} 2mL^3\Omega_m^{2m-1}\ddt \Omega_m \leq 2mL^3 c_{1,m}
\Omega_{m+1}^{m+1}\Omega_m^m 
\EN
which, with some rearranging, becomes 
\EQL{eq:dtOmegam} \dddt{\Omega}_m\leq 
c_{1m,}\left(\frac{\Omega_{m+1}}{\Omega_m}\right)^{m+1} \Omega_m^2\,. \EN
Finally, upon substituting the definition of the $D_m$ and pulling the $\varpi_0$ out,
one gets
\EQL{eq:dtDm}\dddt{D}_m\leq c_{2,m}\varpi_0
\left(\frac{D_{m+1}}{D_m}\right)^{\xi_m} 
D_m^3 \quad{\rm where}\quad \xi_m=\half(4m+1)\,. \EN

Once $\varpi_0=\varpi_\Gamma$ is chosen, then the inviscid $D_m(t)$ can be 
compared for the Euler calculation.  This has been done in same manner 
as in Fig. \ref{fig:DmNSEuler}(left) 
and shows the same ordering as in the Navier-Stokes case\footnote{Note that 
if the rather low frequency of 
$\varpi_\Gamma=\Gamma/L^2$ is replaced by the much larger initial maximum 
of vorticity $\|\omega(t=0)\|_\infty$, this clean ordering is not found.}.
However, in order to include the $m\rightarrow\infty$ limit, a better choice is
to plot $D_m^{-2}(t)$, which for $m=\infty$ gives
$D_\infty^{-2}=\varpi_\Gamma/\|\omega\|_\infty$, where $\|\omega\|_\infty$ 
is the $\sup(|\omega|)$. A simple test for singular behaviour is to compare
$1/\|\omega\|_\infty$ against the power law consistent with the lower bound 
for singular growth of $\|\omega\|_\infty$ allowed by \eqref{eq:BKM}. That is:
\EQL{eq:Dinfm2} \|\omega\|_\infty\sim (T_\infty-t)^{-1}\quad{\rm or}\quad
D_\infty^{-2}=\varpi_\Gamma/\|\omega\|_\infty\sim \varpi_\Gamma(T_\infty-t)\,. \EN
Under this test, the sign of singularity growth would be finding that 
$D_\infty^{-2}\rightarrow0$ linearly.

The $D_m^{-2}(t)$ are plotted in Fig. \ref{fig:DmNSEuler}(right). 
Compared in this way, as $m$ becomes large, the $D_m$ nearly match
$\varpi_\Gamma(T_\infty-t)$ as $t$ increases.
However, the growth of $\|\omega\|_\infty$ appears to tail off of this
behaviour as $t\rightarrow 15$.  So, to claim singular growth, another
independently calculated diagnostic is needed to confirm the trends seen
in Fig. \ref{fig:DmNSEuler}(right). 

If the only diagnostic for singular growth is $\|\omega\|_\infty$, 
then an appropriate secondary diagnostic could be $\alpha=d\log\|\omega\|_\infty/dt$, 
the logarithmic time derivative of $\|\omega\|_\infty$. In principle one should 
determine $\alpha$ from the vortex stretching exactly at the position of 
$\|\omega\|_\infty$. However, to get the stretching at the exact position
of $\|\omega\|_\infty$, which lies between 
the mesh-points in physical space, requires interpolation, 
which can be both difficult and inaccurate. In practice,
the only stretching diagnostic that did not have grid-induced oscillations 
and was near, but not at, the position of $\|\omega\|_\infty$, was to take 
the maximum of the vortex stretching on the perturbation plane \cite{Kerr93}.  

Using the $D_m$ resolves this problem. The trick is to calculate both the $D_m$ 
and their time derivatives $(d/dt)D_m$ at run-time, a simple matter of 
programming compared to the interpolations needed for determining $\alpha$
at $\|\omega\|_\infty$.  Furthermore, Fig. \ref{fig:DmNSEuler}(right) 
shows that as time and $m$ increase, the 
$D_m^{-2}(t)\rightarrow D_\infty^{-2}(t)=\varpi_\Gamma/\|\omega\|_\infty \,.$
Therefore, for large $m$, the secondary diagnostics equivalent to $\alpha$ 
are the logarithmic time derivatives of the $D_m(t)$, which are used below
to define the estimated singular times $T_m(t)$

\subsection*{Numerical analysis using new Euler integrals}

With the added assumption that the $D_{m+1}/D_m$ are always bounded, 
as demonstrated by Fig. \ref{fig:DmNSEuler}(right), 
let us use the bound in \eqref{eq:dtDm} to help us write new 
Euler bounds that can be tested numerically.

\begin{figure}
\begin{center}
\includegraphics[width= 7 cm]{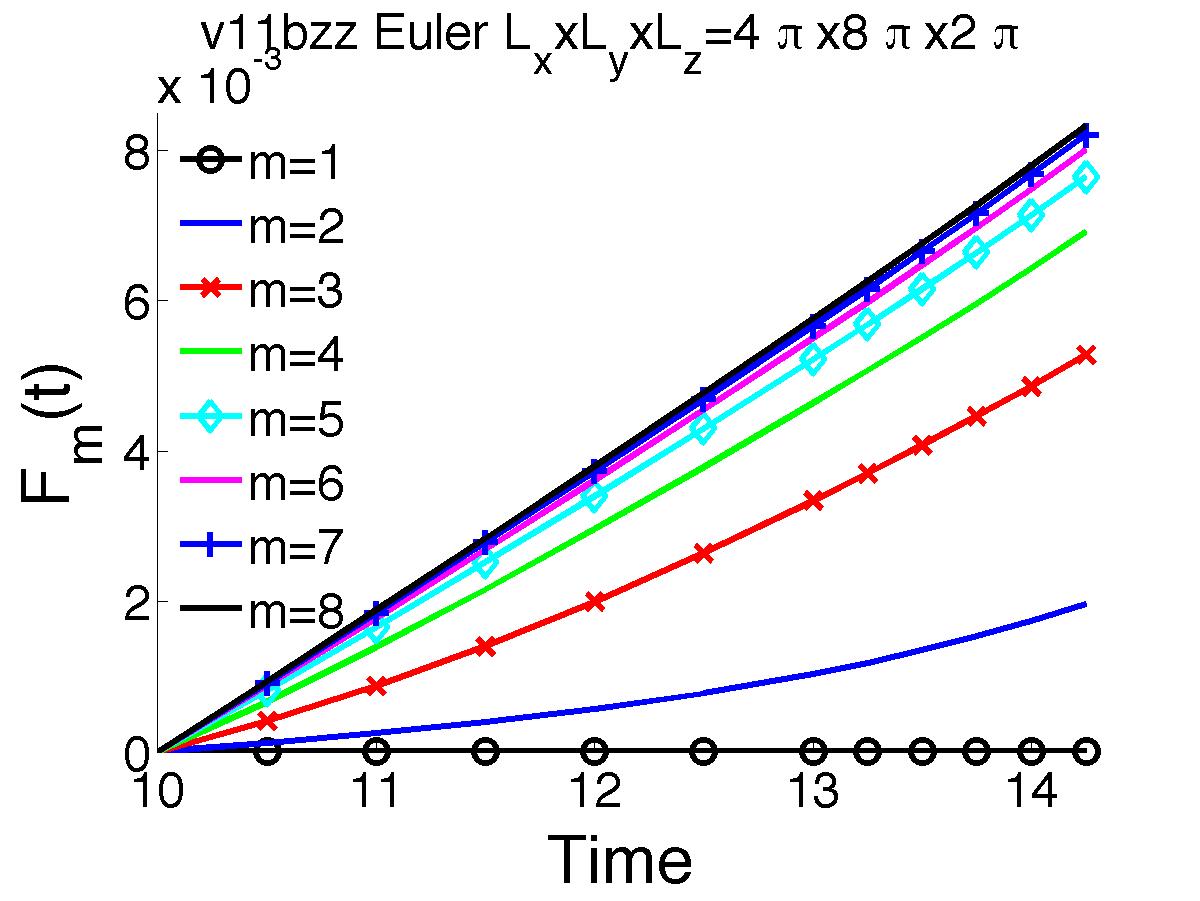}
\hspace*{5mm}
\includegraphics[width= 7 cm]{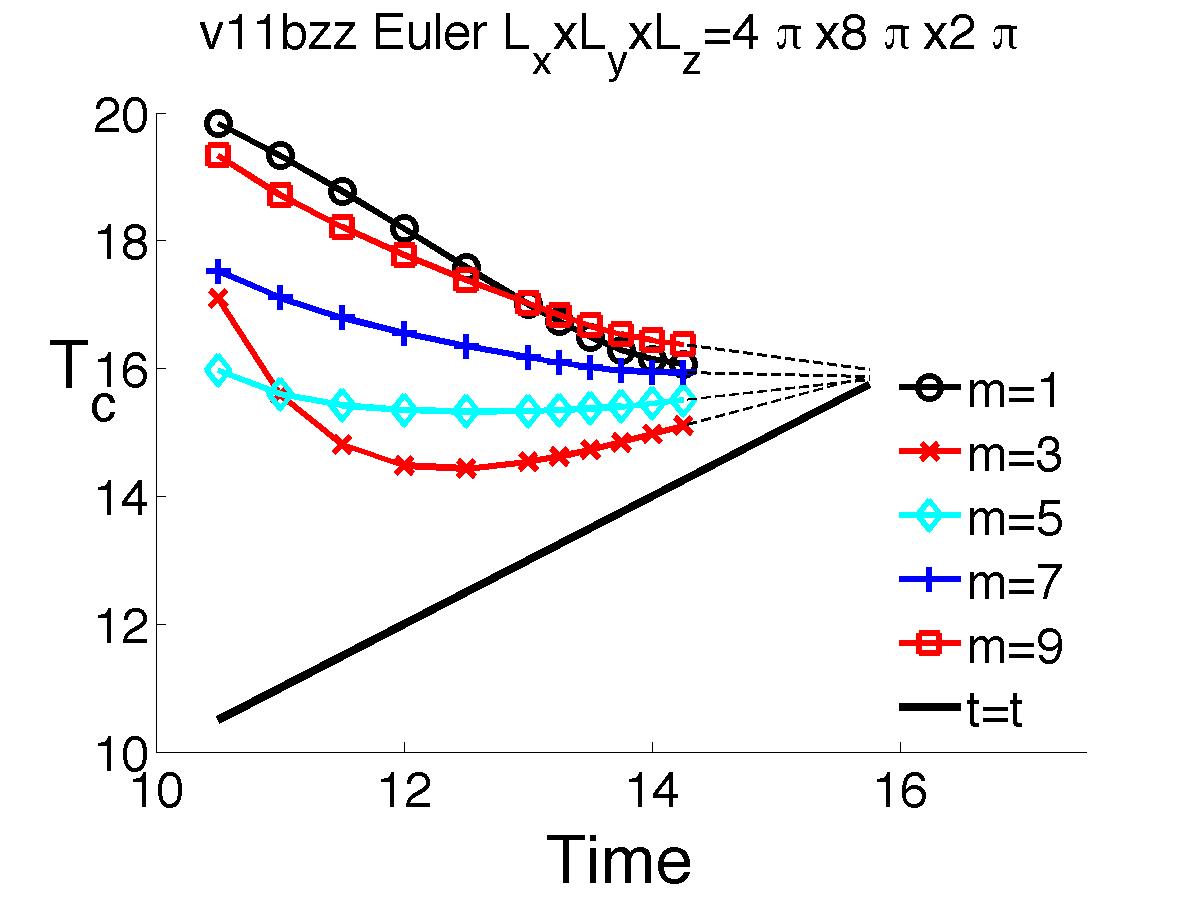}
\vskip6pt
\caption{\label{fig:Dm-2scaling}{\bf Left:}
$\int\varpi_\Gamma(D_{m+1}/D_m)^{\xi_m}dt$ \eqref{eq:dtDmm2}
for an anti-parallel Euler calculation. The observed ordering and linear increase 
as $m\rightarrow\infty$ would permit at least
super-exponential growth of the $D_m$ for both Euler and Navier-Stokes
up to $t=14.5$.
{\bf Right:} Estimated singular time from different $D_m$: 
$T_m(t)=(d\log D_m^2/dt)^{-1}+t$. Only $m$ odd are shown to
reduce clutter and a curve with $t=t$ is added to clarify where
the $T_m(t)$ are heading.  Linear extrapolations to $t=15.75$ 
of the $m>1$ curves, based on the last two values, are shown with 
the dashed lines.  As well as could be expected, these extrapolations 
all appear to be crossing the $t=t$ line at about $T_c\approx 15.8$.
For $t>12$ (and excluding $m=1$), the $T_m(t)$ are ordered. Going from underestimating 
the $T_c$ ($m=3$, 5) to overestimating $T_c$ ($m=7$, 9).
}
\end{center}
\end{figure}

For general $m$, let us begin by rewriting \eqref{eq:dtDm} as
\EQL{eq:dtDmm2} \hspace{-2mm}-\dddt D_m^{-2}\leq 
c_m\varpi_\Gamma\left(\frac{D_{m+1}}{D_m}\right)^{\xi_m}\hspace{-2mm},
\quad\mbox{then using}\quad
F_m(t)=c_m\int_0^t \varpi_\Gamma
\left(\frac{D_{m+1}}{D_m}\right)^{\xi_m}dt\quad\mbox{one gets}\quad
D_m^{-2}\leq c_{2,m} F_m(t)
\,.\EN
This focuses our attention upon the integrals on the right-hand-side, which
are plotted in Fig. \ref{fig:Dm-2scaling}(left).
In this figure, the upper bound, based upon the integral of $D_m^{-2}$, 
grows linearly. If obeyed exactly, this would imply that the solutions 
are singular.  However, since this is only an upper bound, another test is needed.

This final test will be a diagnostic coming from the logarithmic time 
derivatives of the $D_m^{2}$.  Two assumptions are made.  First, that
$D_m^2(t)\sim (T_m-t)^{-\gamma_m}$ and second an assumption on the $\gamma_m$.
Applying $(d\log D_m^2/dt)^{-1}$ to the first
assumption, one gets 
\EQL{eq:Dmm1} (d\log D_m^2/dt)^{-1}=\gamma_m^{-1}(T_m-t)\,. \EN
By applying this assumption to time differences of the $(d\log D_m^2/dt)^{-1}$,
one could get running estimates of both the $\gamma_m$ and the $T_m$.
However, since as $m$ increases the curves generated by the $D_m^2(t)$
are becoming linear in both Fig. \ref{fig:DmNSEuler}(right) 
and Fig. \ref{fig:Dm-2scaling}(left),
that is $\gamma_m\rightarrow 1$, the best way to find running estimates of 
the $T_m(t)$ is to make $\gamma_m\equiv 1$ an added assumption and use
\EQL{eq:Tm} T_m(t)=(d\log D_m^2/dt)^{-1}+t\,. \EN

The result is in Fig. \ref{fig:Dm-2scaling}(right).  For $t\gtrsim 12$,
the estimated $m\geq3$ singular times $T_m(t)$ are beginning to converge.
This is shown more clearly by adding a $t=t$ curve and extending the
computed $T_m$ with linear extensions based on the $T_m$ at the last two
times computed.  If there is a singularity of the Euler equations for
this initial condition, then they should all cross the $t=t$ line at the same 
time.  Which they do.


\begin{table}[h]
\caption{Domains and sequences of meshes used to resolve.}
\begin{tabular*}{\hsize}{@{\extracolsep{\fill}}llcccc@{}}
\hline
Domain & label & Mesh 1 and Mesh 3 & $\Delta t$ & Mesh 2 and Mesh 4 & $\Delta t$ \\
\hline
$3\pi\times3\pi\times2\pi$ & v11a & $512\times256\times512$ & $t=0-12$ & 
$512\times256\times1024$ & $t=8-15$ \\
$3\pi\times3\pi\times2\pi$ & v11a & $1024\times512\times4096$ & $t=10-13.25$ 
& $1024\times512\times2048$ & $t=4-15$ \\
$4\pi\times4\pi\times2\pi$ & v71 & $512\times512\times1024$ & $t=0-12$ & 
$1024\times512\times2048$ & $t=12-13.5$ \\
$4\pi\times4\pi\times2\pi$ & v71 & $1024\times512\times4096$ & $t=13.5-14.25$ \\
$4\pi\times6\pi\times4\pi$ & v11by & $512\times256\times2048$ & $t=0-12$ & 
$1024\times512\times2048$ & $t=8-14$ \\
$4\pi\times4\pi\times4\pi$ & v11bx & $1024\times512\times2048$ & $t=0-14$ & 
$1024\times512\times4096$ & $t=11-14.25$ \\
$4\pi\times8\pi\times4\pi$ & v11bz & $512\times512\times1024$ & $t=0-12$ & 
$1024\times1024\times4096$ & $t=12-13.75$ \\
$4\pi\times8\pi\times2\pi$ & v11bzz & $512\times512\times1024$ & $t=0-12$ & 
$1024\times512\times2048$ & $t=12-13.5$ \\
$4\pi\times8\pi\times2\pi$ & v11bzz & $1024\times512\times4096$ & $t=13.5-14.25$ \\
$4\pi\times16\pi\times2\pi$ & v11g & $512\times1024\times1024$ & $t=0-12$ & 
$512\times1024\times2048$ & $t=12-14.25$ \\
$4\pi\times16\pi\times2\pi$ & v11g & $1024\times2048\times2048$ & $t=10-14$ 
& $1024\times2048\times4096$ & $t=13-14.5$ \\
\hline
\end{tabular*}
\end{table}

\section{Summary}

A new approach to rescaling vorticity moments, the $D_m$, 
has been used for the analysis of new Navier-Stokes and Euler calculations. 
The $D_m$ have the following favourable analytic and numerical 
properties: In mathematical analysis, neighbouring orders can be compared 
using their time derivatives inequalities \cite{Gibbon12b}. In 
numerical analysis, their values, time derivatives and thus their 
logarithmic times derivatives can be determined continuously and compared.

The numerical comparisons have revealed an unexpected hierarchy where the 
lower-order $D_m$ bound the higher-order $D_m$, for all times and for both sets 
of calculations. This ordering of the $D_m$ was unexpected
for two reasons.  First, it is opposite to the required H\"older ordering of the
$\Omega_m$ and second, it is opposite to an ordering that would immediately
imply that the Navier-Stokes equations are regular for all times. 

Furthermore,
the period of strongest growth and alignment of the $D_m(t)$ occurs 
when the normalized enstrophy growth for the Navier-Stokes calculations is strongest.
These observations led to the secondary goal of the Euler calculations. This is 
to use the $D_m$, and their logarithmic growth rates, to determine whether these 
Euler calculations are consistent with the formation a finite-time singularity. 

The new Euler calculations are consistent with the formation a finite-time singularity
in the sense that each of the higher-order moments in Fig. \ref{fig:DmNSEuler} 
show singular trends for a longer period 
than any previous Euler calculation. This includes $D_\infty^2$, 
the rescaled singular maximum of the vorticity $\|\omega\|_\infty$. 
Critical to achieving this extended period of singular Euler growth is 
using vortices that are not subject to internal instabilities and domains 
that are longer than in any earlier work. However, using the new profile and
direction algorithm is not enough.  For the smaller domains listed in Table 1, 
the singular growth saturates early, as in some earlier work \cite{HouLi06}. 
The importance of the longer domain is that it
allows the full effect of the curvature and torsion of the vortices to manifest itself. 
It was not until the length of the domains was $L_y=8\pi$ that 
the extended period of singular growth appeared.

With the new data set and new results, a number of outstanding questions will soon be 
addressed. One is determining the role of the curvature of the vortex lines. 
New analysis shows that the curvature of the vortex lines near $\|\omega\|_\infty$ 
is small, and therefore contributes little to the local $y=0$ vortex stretching. 
This suggests that the stretching is coming from the strong curvature and 
looping seen in Fig. \ref{fig:T0-16} (right) at $t=16$ for $y =\pm5$. 
Even though this structure is forming far from the position of $\|\omega\|_\infty$ 
on the $y=0$ perturbation plane, its effects upon the growth on that plane are
surprisingly strong.  

This success with finding the curvature suggests that are further conditions on 
properties derived from the direction of the vorticity can be determined and tested
against the growth of $\|\omega\|_\infty$ .The new data should be capable of testing 
these proposed constraints and identifying how the position of the 
maximum of vorticity moves with respect to the Lagrangian flow. This relative motion 
is non-zero \cite{KerrBustamante111} and could be important for determining how 
the circulation in the perturbation plane becomes divided into a head 
and flattened tail, where the head is the part of the circulation that 
reconnects if there is viscosity.

\end{document}